
\input PHYZZX.TEX
\font\largemath = cmti10 scaled \magstep2
\textfont4=\largemath
\mathchardef\lS="0453
\footline={\hfill}


\def\Ga{\alpha}
\def\Gb{\beta}
\def\Gc{\chi}
\def\Gd{\delta}
\def\Ge{\epsilon}

\def\Gg{\gamma}
\def\Gh{\eta}

\def\Gp{\pi}
\def\Gq{\theta}

\def\Gt{\tau}

\def\Gw{\omega}

\def\Gy{\psi}

\def\GF{\Phi}

\def\GY{\Psi}

\def\Pd{\partial}
\def\Vp{\varphi}


\leftline{ hep-th/9304058}
\rightline{ CHIBA-EP-68 }

\titlepage
\title{ Superfield Formalism of Stochastic Quantization Method
 with Field-Dependent Kernels}

\author{Kenji Ikegami}

\address{Graduate School of Science and Technology,  \break
                 Chiba University, \break
        1-33 ~Yayoi-cho,  ~Inage-ku, ~Chiba 263, JAPAN }%

\baselineskip = 12pt

\abstract{I consider a Langevin equation with field-dependent
kernels and investigate supersymmetry of the stochastic generating
functional constructed from the Langevin equation. Moreover I
describe the stochastic generating functional in terms of
a superfield. In the superfield formalism, it becomes clear that
the stochastic quantization method with
the field-dependent kernel is equivalent to
the path-integral quantization method.}%

\baselineskip = 18pt
%
\REF \parisi{G.Parisi and Y.Wu, Sci.Sin. {\bf 24}
(1981)483.}
\REF \damgaad{For a review, P.H.Damgaad and H.H\"uffel,
Phys.Rep.{\bf 152}(1987)227.}
\REF \ghost{M.Namiki, I.Ohba, K.Okano and Y.Yamanaka,
Prog. Theor. Phys. {\bf 69}\hfill\break (1983)1580.}
\REF \nakghost{A.Nakamura, Prog.Theor.Phys.
{\bf 86}(1991)925.}
\REF \regularization{For example, J.T.Breit, S.Gupta and A.Zaks,
Nucl.Phys.B233 (1984)61;\hfill\break
J.Alfaro, Nucl.Phys.B253(1985)464.}
\REF \fermi{U.Kakudo, Y.Taguchi, A.Tanaka and K.Yamanoto,
Prog.Theor.Phys.{\bf 69}\hfill\break
(1983)1225;\hfill\break
T.Fukai, H.Nakazato,I.Ohba,K.Okano and Y.Yamanaka, Prog.Theor.Phys.
{\bf 69}(1983)1600.}
\REF \gravity{H.Rumpf, Phys.Rev.D33(1986)942.}
\REF \ikemocyos{K.Ikegami, R.Mochizuki and K.Yoshida,
Prog.Theor.Phys.{\bf 89}(1993)197.}
\REF \tanaka{S.Tanaka, M.Namiki, I.Ohba, M.Mizutani, N.Komoike and
M.Kanenaga, \hfill\break Phys.Lett.B288(1992)129.}
\REF \ps{G.Parisi and N.Sourlas, Phys.Rev.Lett.43(1979)744;
\hfill\break Nucl.Phys.B206(1983)321.}
\REF \mockernel{R.Mochizuki, Prog.Theor.Phys. {\bf 85}(1991)407}
\REF \FP{A.D.Fokker, Ann.Physik 43(1914)810;\hfill\break
M.Plack, Sitzber.Pr.Akad.Wiss.(1917)324.}
\REF \cardy{J.L.Cardy, Phys.Lett.B125(1983)470.}
\REF \kugo{T.Kugo and M.G.Mitchard, Prog.Theor.Phys.
{\bf 83}(1990)134.}
\REF \sakita{B.Sakita, in Proc.Johns Hopkins Workshop 7,
eds.Domokos and Kovesi-\break Komokos (World Scientific, Singapore,
1983);\hfill \break
K.Ishikawa, Nucl.Phys.{\bf B241}(1984)589 \hfill}
\REF \stra{R.L.Stratonovich, Conditional Markov Processes and
Their Application to the Theory of Optimal Control(Elsevier,
NewYork, 1968)}
\REF \super{E.Egorian and S.Kalitzin, Phys.Lett.B129(1983)320;
\hfill\break
R.Kirscher, Phys.Lett.B139(1984)180;\hfill\break
E.Gozzi, Phys.Lett.B139(1984)183.}

%
\footline={\hfill-- \folio\ --\hfill}
\pagenumber = 1


\chapter{Introduction}
 The stochastic quantization method (SQM) was first
proposed by Parisi and
Wu as an alternative quantization method in 1981.
\refmark\parisi \refmark\damgaad
SQM can be applied to gauge theories without the
gauge fixing procedure, i.e. without Faddeev-Popov ghost
fields. Instead of introducing ghost field, the method
produces the same contribution as the path-integral quantization
method (PIQM). This fact was already confirmed perturbatively for
Yang-Mills fields\refmark\ghost
and for non-Abelian anti-symmetric tensor fields.
\refmark\nakghost

SQM has a powerful tool, `` kernel'', which, among others,
gives new regularization schemes.\refmark\regularization
Kernel is
also introduced for system including massless fermion.
\refmark\fermi
Moreover, the ``field-dependent'' kernel is introduced for
system including \break
graviton,\refmark\gravity
system with spontaneously broken symmetry,
\refmark\ikemocyos
and bottomless \break systems.\refmark\tanaka
On the other hand, it is well known that theories quantized
stochastically display
supersymmetry\refmark\ps and can be described in superfield formalism.
So my question is whether SQM
with field-dependent kernel has supersymmetry or
not. While Ref.[\mockernel]
showed that stochastic action with field-dependent kernel
had a supersymmetry, the action cannot be
described in superfield formalism. Besides, the stochastic
action is correct only when the field-dependent kernel is
a metric included in classical action. In this paper,
I show that the generating functional is invariant under
two independent super-transformations and can be described
in terms of superfield for boson and fermion systems
in general.

 SQM without kernel is equivalent to PIQM and proof
of the equivalence is given with the help of Fokker-Planck
equation\refmark\FP or superfield formalism.
\refmark\cardy\refmark\kugo
The equivalence is, however, not given yet in
superfield formalism in case Langevin equation
has field-dependent kernel.
I remark the latter equivalence in this paper.

 This paper is organized as follows.
Supersymmetric generating functional for boson system is given
in section 2 and for fermion system in
section 3. In section 4,
the equivalence of SQM with field-dependent kernel
to PIQM is remarked and summary is given.

\chapter{Supersymmetric generating functional for boson}

I take up a system with variables $q(x)$ and
the classical action $S(q)$ in n-dimensional space-time.
To quantize the system,
I give a Langevin equation with field-dependent kernel
$K(q)$
$$
\Pd_t q(x,t) \equiv \dot{q}(x,t)
 = -X(x,t) + R(q(x,t)) \Gh (x,t), \eqno(1)
$$
$$
   X \equiv K(q(x,t)){\Gd S \over \Gd q(x,t)}
    -R(q(x,t)){\Gd R(q(x,t)) \over \Gd q(x,t)},\ \ \ \
  K(q(x,t)) = R^2(q(x,t)),
$$
where $\Gh$
is a white noise field defined by the following correlation
$$
\eqalign{ \langle \Gh(x,t) &\Gh(y,s) \rangle_{\Gh}
                    = 2\Gd^n (x-y) \Gd(t-s),\cr
 \langle f(\Gh) \rangle_{\Gh} &\equiv
\int D\Gh \ f(\Gh)\
   exp\{-\smallint d^n x dt {1\over 4}\Gh^2(x,t)\}.}\eqno(2)
$$
$f(\Gh)$ is an arbitrary
function of $\Gh$. In this paper, only Stratonovich type
calculus\refmark\stra is used which allows the Leibnitz rule with
respect to stochastic time derivative.
Now, let me introduce the stochastic generating functional
$$
Z[j] = \langle e^{\int d^n x dt \ q_{\Gh}(x,t) j(x,t)}\rangle_{\Gh}
     = \int D\Gh \ e^{-\int d^nx dt\{ {1\over 4}\Gh^2(x,t)
                  - q_{\Gh}(x,t) j(x,t)\} }, \eqno(3)
$$
where $q_{\Gh}$ is solution
of eq.(1). Inserting the right-hand side of
$$
1=\int Dq\ \Gd (R^{-1}(\dot{q}+X)-\Gh){\Gd \Gh \over \Gd q},
$$
I get
$$
\eqalign{Z[j]
= \int &Dq
      Det\lbrack{\Pd \over \Pd q}\lbrace R^{-1}(q)
         (\dot{q}+X(q))\rbrace\rbrack
           \cr
& \times  exp\lbrack -\int d^n x dt
          \{ {1\over 4}(\dot{q}+X(q))K^{-1}(\dot{q}+X(q))
          - j(x,t)q(x,t)\} \rbrack
   ,\cr
= \int &Dq D\Gw D\overline{\Gw} Dp \
      exp\big[ -\int d^n x dt [ pK^{-1}p
       -ip(\dot{q} +X(q))
          \cr
&\ \ \ \ \ \ \ \ \ \ \ \ \ \ \ \ \ \ \ \ \ \ \
           -\overline{\Gw} R {\Pd \over \Pd q}
        \lbrace R^{-1}(\dot{q}+ X(q))\rbrace
        \Gw  -j(x,t)q(x,t) ]\big],}\eqno(4)
$$
where $\Gw,\overline{\Gw},p$ are auxiliary fields.
This expression is rather complicated and it is difficult to
recognize in eq.(4) whether $Z[0]$ has supersymmetry or not.
In fact, the stochastic action appearing in eq.(4)
$$
\eqalign{S_{SQM}(q,\Gw,\overline{\Gw},p) \equiv
     \int d^nx d\tau \lbrace pK^{-1}p& -ip(\dot{q}
        +K{\Gd S \over \Gd q}- R{\Gd  R\over \Gd q})\cr
        &-\overline{\Gw} R {\Pd \over \Pd q}
          ( R^{-1}\dot{q}
       +  R{\Gd S \over \Gd q}-{\Gd  R\over \Gd q})
        \Gw \rbrace,}\eqno(5)
$$
is not invariant under the supersymmetric transformation
$$
\Gd q = \overline{\Ge}\Gw + \overline{\Gw}\Ge,
\ \ \ \ \ \Gd \overline{\Gw} = -i\overline{\Ge}p,
\ \ \ \ \ \Gd \Gw = -i\Ge p -\Ge \dot{q},
\ \ \ \ \ \Gd p = i\dot{\overline{\Gw}}\Ge, \eqno(6)
$$
which makes the stochastic action without kernel invariant.
\refmark\ps\refmark\super
Here $\Ge,\overline{\Ge}$ are infinitesimal anticommuting
constant parameters.
The change of variables
$$
q'=\int dq R^{-1}(q), \ \ \ p'= R(q)p,\ \ \
\overline{\Gw}'=\overline{\Gw} R(q),
\ \ \ \Gw'= R^{-1}(q)\Gw,\eqno(7)
$$
leads to
$$
\eqalign{Z[j] = \int &Dq' D\Gw' D\overline{\Gw}' Dp' exp\lbrack
    -\smallint dx dt \lbrace p'^2 -ip'(\dot{q}'
        +{\Gd S \over \Gd q'})\cr
      & -\overline{\Gw}' {\Pd \over \Pd q'}
          (\dot{q}'+ {\Gd S \over \Gd q'}) \Gw'
+ip' R^{-1}{\Gd  R\over \Gd q'}
       +\overline{\Gw}'{\Pd \over \Pd q'}
          ( R^{-1}{\Gd  R\over \Gd q'})\Gw'-j(x,t)q(q')
             \rbrack,}\eqno(8)
$$
where I assume that the first relation in eq.(7) can be
solved for $K$ in terms of $q$. $Z[j]$ can then
be rewritten in terms of a superfield $\GF'$ as
$$
\eqalign{Z[j]= \int D\GF' exp\lbrack
              -\int d^2\Gq d\Gt d^n x\lbrace &
       \overline{D}_{\Gq}\GF'D_{\overline{\Gq}}\GF'+ L(q(\GF'))
        \cr
        &-\Gd^n(0)\ln R(q(\GF'))-jq(q')\rbrace,}\eqno(9)
$$
$$
\eqalign{\GF'\equiv q' +
\overline{\Gq}\Gw' + \overline{\Gw}'\Gq
    -i\overline{\Gq}\Gq p',
\ \ &D\GF'\equiv Dq'D\Gw' D\overline{\Gw}' Dp',
               \cr
\ \ \overline{D}_{\Gq}\equiv \Pd_{\Gq},
\ \ D_{\overline{\Gq}} \equiv &
\Pd_{\overline{\Gq}}-\Gq\Pd_{\tau},}\eqno(10)
$$
where $\Gq,\overline{\Gq}$ are anticommuting superspace
coordinates,
$L(q)$ is Lagrangian density $\int d^nx L =S$, and
$q(\GF')\equiv q(q'))|_{q'=\GF'}$. Now it is obvious that
the stochastic action
$$
\eqalign{& S'_{SQM}(\GF')=S_{SQM}(q,\Gw,\overline{\Gw},p)
\cr
& =\int d^2\Gq d\tau d^nx\lbrace
              \overline{D}_{\Gq}\GF' D_{\overline{\Gq}}\GF'
           +L(q(\GF'))-\Gd^n(0)\ln R(q(\GF'))\rbrace,}\eqno(11)
$$
is invariant by operation with supercharges $Q,\overline{Q}$
$$
Q\equiv \Pd_{\overline{\Gq}},
\ \ \ \ \overline{Q}\equiv \Pd_{\Gq} + \overline{\Gq}\Pd_{\tau},
\ \ \ \ \lbrace Q,D_{\overline{\Gq}}\rbrace=
        \lbrace \overline{Q}, D_{\overline{\Gq}}\rbrace=
        \lbrace Q,\overline{D}_{\Gq}\rbrace=
        \lbrace \overline{Q},\overline{D}_{\Gq} \rbrace=0,\eqno(12)
$$
or equivalently under the supertransformation
$$
\Gd q'= \overline{\Ge}\Gw'+\overline{\Gw}'\Ge,
\ \ \ \ \ \Gd\overline{\Gw}'=-i\overline{\Ge}p',
\ \ \ \ \ \Gd\Gw'=-i\Ge p'-\Ge\dot{q}',
\ \ \ \ \ \Gd p'= i\dot{\overline{\Gw}}'\Ge.\eqno(13)
$$
In terms of original variables $q,\Gw,\overline{\Gw},p$, the
transformation can be expressed as
$$
\eqalign{&\Gd q= \overline{\Ge}\Gw +\overline{\Gw}K(q)\Ge,
\ \ \ \Gd\overline{\Gw}=-i\overline{\Ge}p'-\overline{\Ge}\Gw
             \overline{\Gw}{\Pd  R(q)\over \Pd q} R^{-1}(q)
           ,\cr
&     \Gd\Gw=-i\Ge K(q)p -\Ge\dot{q}
           + \overline{\Gw} \Ge {\Pd  R(q)\over \Pd q}R(q)\Gw,
\cr
&\Gd p= i\dot{\overline{\Gw}}\Ge
            +i\overline{\Gw}{\Pd R(q)\over \Pd q}R^{-1}\dot{q}\Ge
            -R^{-1}(q){\Pd R(q)\over \Pd q}\overline{\Ge}\Gw p
            -\overline{\Gw}\Ge {\Pd R(q)\over \Pd q}R(q) p
.}\eqno(14)
$$
Thus the stochastic action with field-dependent kernel
is invariant, for any boson system, under
the super-transformation. With the help of the generating
functional (9), it is shown
that the Green functions in SQM are equivalent to those
in PIQM as will be remarked in section 4.

\chapter{Fermion case}
 Next, I show the supersymmetry for the fermion system which
is quantized stochastically.
In general, as kernel $K(\Gy,\overline{\Gy})$
for fermion field includes
Dirac matrices $\Gg_{\mu}$, $\sqrt{K}$ does not always exist.
So I start with the Langevin equation
as \refmark\sakita
$$
\eqalign{&\dot{\Gy}_{\Ga}(x,t)=
-X_{\Gy}
+\Gh_{1\Ga} + {1\over 2}K_{\Ga \Gb}\Gh_{2\Gb}
,\cr
&\overline{\Gy}_{\Ga}(x,t)=
-X_{\overline{\Gy}}
+{1\over 2}\overline{\Gh}_{1\Gb}K_{\Ga \Gb}
+\overline{\Gh}_{2\Ga},}\eqno(15)
$$
$$
X_{\Gy}\equiv K_{\Ga \Gb}{\Gd S \over \Gd \overline{\Gy}_{\Gb}}
-{1\over 2}{\Gd K_{\Ga \Gb} \over \Gd\overline{\Gy}_{\Gb}},\ \ \
X_{\overline{\Gy}}\equiv -{\Gd S \over \Gd \Gy_{\Gb}}K_{\Gb \Ga}
+ {1\over 2}{\Gd K_{\Gb \Ga} \over \Gy_{\Gb}},\eqno(16)
$$
where $\Gd /\Gd \Gy, \Gd/\Gd \overline{\Gy}$ are
left derivatives and
$\Gh_1,\Gh_2,\overline{\Gh}_1,\overline{\Gh}_2$ are
anticommuting white noise fields defined as
$$
\langle \Gh_{1\Ga}(x,t) \overline{\Gh}_{1\Gb}(x',t')\rangle
=\langle \Gh_{2\Ga}(x,t) \overline{\Gh}_{2\Gb}(x',t')\rangle
=2\Gd_{\Ga \Gb}\Gd^n(x-x')\Gd(t-t'),\eqno(17)
$$
$$
\langle
f(\Gh_1,\Gh_2,\overline{\Gh}_1,\overline{\Gh}_2)
\rangle
\equiv \int D\overline{\Gh}_1D\Gh_1 D\overline{\Gh}_2 D\Gh_2
\
f(\Gh_1,\Gh_2,\overline{\Gh}_1,\overline{\Gh}_2)
\
e^{-{1 \over 2}\int d^nx d\tau(
\overline{\Gh}_1\Gh_1 + \overline{\Gh}_2\Gh_2)}.\eqno(18)
$$
The stochastic generating functional
for fermi field is defined as
$$
Z[j,\overline{j}]=
\int D\overline{\Gh}_1D\Gh_1 D\overline{\Gh}_2 D\Gh_2
e^{ -\int d^nx dt\{
{1\over 2}(\overline{\Gh}_1\Gh_1 +\overline{\Gh}_2\Gh_2)
-(\overline{j}\Gy_{\Gh}+\overline{\Gy}_{\Gh}j) \},}\eqno(19)
$$
where $\Gy_{\Gh},\overline{\Gy}_{\Gh}$ are solutions of eq.(15).
The change of variables $(\Gh_1,\overline{\Gh_1},
\Gh_2,\overline{\Gh_2})\rightarrow (\Gy,\overline{\Gy},
\Gh'_2 \equiv K\Gh_2,
\overline{\Gh}'_2 \equiv \overline{\Gh}_2K^{-1})$ leads to
$$
\eqalign{Z[j,\overline{j}]=
\int D&\overline{\Gy} D\Gy D\overline{\Gh}'_2
D\Gh'_2
D\overline{\Vp}_1 D\Vp_1 D\overline{\Vp}_2 D\Vp_2\
Det(K)\ exp\Bigl[ -\int d^nx d\tau\{
\overline{\Gh}'_2 \Gh'_2
\cr
&-\overline{\Gh}'_2 (\dot{\Gy}+X_{\Gy})
-{1\over 2}(\dot{\overline{\Gy}}+X_{\overline{\Gy}})
K^{-1} \Gh'_2
+(\dot{\overline{\Gy}}+X_{\overline{\Gy}})K^{-1}
(\dot{\Gy} +X_{\Gy})
\cr
&\ \ \ \ \
-\left( \overline{\Vp}_1\ \overline{\Vp}_2 \right)
\Bigl( J \Bigr)
\left( \eqalign{&\Vp_1 \cr &\Vp_2}\right)
-(\overline{j}\Gy_{\Gh}+\overline{\Gy}_{\Gh}j)\} \Bigr]
,}\eqno(20)
$$
$$
\left( J\right)_{\Ga \Gb} \equiv\pmatrix{
-{\Pd \over \Pd\Gy_{\Gb}}(\dot{\Gy}_{\Ga}+X_{\Gy})_{\Ga}
 &\hfil\hfil
-{\Pd \over \Pd\overline{\Gy}_{\Gb}}(\dot{\Gy} +X_{\Gy})_{\Ga}
\hfill\cr
\hfil {\Pd \over \Pd\Gy_{\Gb}}\{ (\dot{\overline{\Gy}}
+X_{\overline{\Gy}})_{\Gc}K^{-1}_{\Gc\Gg}\}K_{\Gg\Ga}
&\hfil
{\Pd \over \Pd\overline{\Gy}_{\Gb}}
\{ (\dot{\overline{\Gy}}+X_{\overline{\Gy}})_{\Gc}
K^{-1}_{\Gc\Gg}\} K_{\Gg\Ga}\hfill\cr
},
$$
where $\Vp,\overline{\Vp}$ are auxiliary fields.
After the integration over $\Gh'_2,\overline{\Gh}'_2$,
$$
\eqalign{Z[j,\overline{j}] =
\int D\overline{\Gp} D\Gp D\overline{\Vp}_2
D\Vp_2 D\overline{\Vp}_1 D\Vp_1 D\overline{\Gy} D\Gy & \
\cr
\times exp\Big[ -\int d^nx d\tau \{2\overline{\Gp}K\Gp
-i\overline{\Gp}(\dot{\Gy}+& X_{\Gy})
-i(\dot{\overline{\Gy}}+ X_{\overline{\Gy}})\Gp
\cr
-\left( \overline{\Vp}_1\ \overline{\Vp}_2\right)
(J)\left(\eqalign{&\Vp_1 \cr &\Vp_2 }\right)
-&(\overline{j}\Gy_{\Gh}+\overline{\Gy}_{\Gh}j)\} \Big]
,}\eqno(21)
$$
where $\Gp,\overline{\Gp}$ are auxiliary fields.
As in section 2, the change of variables
$$
\matrix{\Gy'=\Gy,\hfil &\Vp'_1=\Vp_1,
\hfil &\overline{\Vp}'_1=\overline{\Vp}_1,
\hfil &\overline{\Gp}'=\overline{\Gp},
\cr
\Gd\overline{\Gy}'=\Gd\overline{\Gy}K^{-1},
\hfil &\Vp'_2=\Vp_2K^{-1},
\hfil &\overline{\Vp}'_2=K\overline{\Vp}_2,
\hfil &\Gp'=K\Gp,}\eqno(22)
$$
leads to the generating functional $Z[j,\overline{j}]$ written in
terms of superfield
$$
\eqalign{Z[j,\overline{j}]=\int D\overline{\GY}' D\GY' \
exp\Big[-\int & d^nx d^2\Gq d\tau \{
     D_{\overline{\Gq}}\overline{\GY}'  \overline{D}_{\Gq}\GY'
     -\overline{D}_{\Gq}\ \overline{\GY}' D_{\overline{\Gq}}\GY'
     \cr
     & +L(\GY',\overline{\GY}')
     -{1\over 2}\Gd^n(0)\ln \ detK(\GY',\overline{\GY}' )
   \} \Big],}\eqno(23)
$$
$$
\GY'\equiv \Gy'+\overline{\Gq} \Vp'_1
+ \overline{\Vp}'_2\Gq -i \overline{\Gq}\Gq \Gp',\ \ \ \
\overline{\GY}'\equiv \overline{\Gy}'+\overline{\Gq}\Vp'_2
+ \overline{\Vp}'_1\Gq -i \overline{\Gq}\Gq\overline{\Gp}'
,\eqno(24)
$$
where $L$ is Lagrangian density and $\overline{\Gy}'$
is defined from
${\Pd\overline{\Gy}'\over\Gd\overline{\Gy}}=K^{-1}$.
The supersymmetry transformation in terms of
original fields is
$$
\eqalign{&\Gd \Gy_{\Ga}=\overline{\Ge}\Vp_{1\Ga}
 +K_{\Ga \Gb}\overline{\Vp}_{2\Gb}\Ge,\
 \Gd \Vp_{1\Ga}=-i\Ge K_{\Ga \Gb}\Gp_{\Gb}-\Ge \dot{\Gy}_{\Ga},\
 \Gd \overline{\Vp}_{1\Ga}=
              i\overline{\Ge}\ \overline{\Gp}_{\Ga},\
 \Gd \overline{\Gp}_{\Ga}=i\Ge \dot{\overline{\Vp}}_{\Ga},
\cr
&\Gd \overline{\Gy}_{\Ga}=
  \overline{\Ge}\Vp_{2\Ga}+\overline{\Vp}_{1\Gb}K_{\Gb\Ga}\Ge,
\cr
& \Gd \Vp_{2\Ga}=-i\Ge \overline{\Gp}_{\Gb}K_{\Gb\Ga}
   -\Ge\dot{\overline{\Gy}}_{\Ga}
   -\Vp_{2\Gg}(\overline{\Ge}\Vp_{1}
         +K\overline{\Vp}_{2}\Ge)_{\Gd}
    {\Pd K^{-1}_{\Gg\Gb} \over \Pd \Gy_{\Gd}}K_{\Gb\Ga}
  \cr
&\ \ \ \ \ \ \ \ \ \ \ \ \ \ \ \ \ \ \ \ \ \ \ \ \ \ \ \ \ \ \ \
\ \ \ \ \ \ \ \ \ \ \ \ \ \ \ \ \ \ \
     -\Vp_{2\Gg}(\overline{\Ge}\Vp_{2}
         +\overline{\Vp}_{1}K\Ge)_{\Gd}
    {\Pd K^{-1}_{\Gg\Gb} \over \Pd \overline{\Gy}_{\Gd}}
                                         K_{\Gb\Ga},
\cr
& \Gd \overline{\Vp}_{2\Ga}=i\overline{\Ge}\Gp_{\Ga}
   -(\overline{\Ge}\Vp_{1}+K\overline{\Vp}_{2}\Ge)_{\Gd}
       K^{-1}_{\Ga\Gb} {\Pd K_{\Gb\Gg} \over \Pd \Gy_{\Gd}}
             \overline{\Vp}_{2\Gg}
   -(\overline{\Ge}\Vp_2+\overline{\Vp}_1K\Ge)_{\Gd}K^{-1}_{\Ga\Gb}
      {\Pd K_{\Gb\Gg} \over \Pd \overline{\Gy}_{\Gd}}
           \overline{\Vp}_{2\Gg},
\cr
& \Gd\Gp=i\Ge\dot{\overline{\Vp}}_{2\Ga}
   +i\Ge \dot{\Gy}_{\Gd}K^{-1}_{\Ga\Gb}
          {\Pd K_{\Gb\Gg}\over\Pd\Gy_{\Gd}}\Vp_{2\Gg}
   -(\overline{\Ge}\Vp_1 + \overline{\Vp}_2K\Ge)_{\Gd}K^{-1}_{\Ga\Gb}
          {\Pd K_{\Gb\Gg} \over \Pd \Gy_{\Gd}}\Gp_{\Gg}
\cr
&\ \ \ \ \ \ \ \ \ \ \ \ \ \ \ \ \ \ \ \ \ \ \ \ \ \ \ \ \ \ \ \
\ \ \ \ \ \ \ \ \ \ \ \ \ \ \ \ \ \ \ \ \
   -(\overline{\Ge}\Vp_2 + \overline{\Vp}_1K\Ge)_{\Gd}K^{-1}_{\Ga\Gb}
          {\Pd K_{\Gb\Gg}\over\Pd\overline{\Gy}_{\Gd}}\Gp_{\Gg}.
}\eqno(25)
$$
which is the supersymmetry for $S_{SQM}$.

\chapter{Summary}
 I showed that the stochastic generating functional for fermion
or boson system, which is constructed from the Langevin
equation with field-dependent kernel, has
supersymmetry and can be described in terms of superfield.

 Further I remarked that SQM with field-dependent kernel
is equivalent to PIQM. The generating functional
written in terms of superfield is identical to that of
Ref.[\kugo] which is constructed from the Langevin equation
without field-dependent kernel with the replacement of
$\GF$ with $\GF'$ and $L(\GF)$ with $L(q(\GF'))-\ln(R(q(\GF'))$.
In Ref.[\kugo] the equivalence of SQM without
kernel to PIQM was proved when ${\Pd^2 L\over \Pd q^2}$ is positive.
So the equivalence of SQM with field-dependent kernel to PIQM
is proved in the same way if
$$
{\Pd^2 (L-\Gd^n(0)\ln R)\over \Pd q^2 }
$$
is positive.

\chapter{Acknowledgments}
 Auther thanks Prof.S.Kawasaki, Drs.R.Mochizuki and A.Nakamura
for valuable discussion and comments.

\vfill\eject

\refout

\vfill\eject

\bye